\documentclass[a4paper]{article}

\usepackage{INTERSPEECH2020}
\usepackage{amsmath,graphicx}

\usepackage{centernot}
\usepackage[ruled,vlined]{algorithm2e}
\SetAlFnt{\scriptsize}

\usepackage{wrapfig}
\usepackage{setspace}
\usepackage{cleveref}

\usepackage{multirow}

\makeatletter
\renewcommand{\section}{\@startsection
  {section}%
  {1}%
  {}%
  {-0.5\baselineskip}%
  {0.2\baselineskip}%
  {}}%

\renewcommand{\subsection}{\@startsection
  {subsection}%
  {2}%
  {}%
  {-0.1\baselineskip}%
  {0.1\baselineskip}%
  {}}%

\renewcommand{\subsubsection}{\@startsection
  {subsubsection}%
  {3}%
  {}%
  {-0.2\baselineskip}%
  {0.2\baselineskip}%
  {}}%

\g@addto@macro\normalsize{%
  \setlength\abovedisplayskip{5pt plus 2pt minus 2pt}
  \setlength\belowdisplayskip{5pt plus 2pt minus 2pt}
  \setlength\abovedisplayshortskip{4pt plus 2pt minus 2pt}
  \setlength\belowdisplayshortskip{4pt plus 2pt minus 2pt}
}

\makeatother

\Crefname{equation}{Eq.}{Eqs.}
\Crefname{figure}{Fig.}{Figs.}
\Crefname{tabular}{Tab.}{Tabs.}

\newcommand{\pluseq}{\mathrel{+}=}
\newcommand{\prodeq}{\mathrel{\cdot}=}

\DeclareMathOperator*{\argmax}{arg\,max}

\newcommand\numberthis{\addtocounter{equation}{1}\tag{\theequation}}

\title{Robust Beam Search for Encoder-Decoder Attention Based \\ Speech Recognition without Length Bias}
\name{Wei Zhou, Ralf Schl\"uter, Hermann Ney}
\address{
  Human Language Technology and Pattern Recognition, Computer Science Department,\\
  RWTH Aachen University, 52074 Aachen, Germany\\
  AppTek GmbH, 52062 Aachen, Germany}
\email{\{zhou, schlueter, ney\}@cs.rwth-aachen.de}

\begin{document}

\maketitle
\begin{abstract}
As one popular modeling approach for end-to-end speech recognition, attention-based encoder-decoder models are known to suffer the length bias and corresponding beam problem. Different approaches have been applied in simple beam search to ease the problem, most of which are heuristic-based and require considerable tuning. We show that heuristics are not proper modeling refinement, which results in severe performance degradation with largely increased beam sizes. We propose a novel beam search derived from reinterpreting the sequence posterior with an explicit length modeling. By applying the reinterpreted probability together with beam pruning, the obtained final probability leads to a robust model modification, which allows reliable comparison among output sequences of different lengths. Experimental verification on the LibriSpeech corpus shows that the proposed approach solves the length bias problem without heuristics or additional tuning effort. It provides robust decision making and consistently good performance under both small and very large beam sizes. Compared with the best results of the heuristic baseline, the proposed approach achieves the same WER on the `clean' sets and 4\% relative improvement on the `other' sets. We also show that it is more efficient with the additional derived early stopping criterion.
\end{abstract}

\vspace{-0.5mm}
\section{Introduction \& Related Work}
So called ``end-to-end" speech recognition enables the direct mapping of acoustic feature sequences to sub-word or word sequences. One of the most successful end-to-end approaches is the attention-based encoder-decoder model \cite{attentionNMT2015}, which has achieved promising results in speech recognition \cite{LAS, specaug, Zeyer2019ASRU, Zoltan2020swb}. 
For attention-based encoder-decoder systems without monotonic constraints, there is generally no explicit time or positional information in the output sequences w.r.t the input sequences. Such systems usually apply label-synchronous search for decoding, where mostly a sequence end label is used for termination. Simple beam search is used for most end-to-end systems, where only an absolute beam size limit controls the complete search procedure.

Encoder-decoder models are known to suffer the length bias problem due to the locally normalized training objective \cite{lengthBias2016}. In short, models can produce higher sequence posterior for much shorter output sequences than for the correct ones.
This behavior becomes more obvious with larger beam sizes in beam search, which leads to the beam problem. Reasonable performance is only achieved with very small beam sizes, where search errors are adopted to avoid model errors \cite{NMTsearchError2019}. Such issues are observed in many applications such as speech recognition \cite{Chorowski2017coverage, EOS} and neural machine translation (NMT) \cite{legntNMT, MurrayC18WMT, googleNMT2016}. While an additional length model may solve the problem, it is often ignored or considered to be implicitly learned with existing models.

Instead, many different approaches have been applied in simple beam search to ease the problem. The general goal is to prevent too short output sequences and to allow reliable comparison among output sequences of different lengths. In decoding, scores are commonly used by taking logarithm of probabilities. One straightforward and widely-used approach is the length normalization \cite{LengthNorm, MurrayC18WMT}, which divides the score of a sequence by its length. Another common approach is the end-of-sentence (EOS) threshold \cite{EOS, Chorowski2017coverage}, which allows a sequence end label to appear only if its score is better than the current best non-end one multiplied by a predefined factor. \cite{Zeyer2019ASRU} combined these two approaches and obtained good results with beam size 64, which we will use as our baseline. Another approach is a length reward term added to the score of a sequence based on its length \cite{CtcLenReward, Bahdanau2016e2e, He2016wordreward}, where the scaling value requires careful tuning on each data set. A more sophisticated but less heuristic-based approach is the coverage term \cite{Chorowski2017coverage, coverageNMT}, which is added to the score based on the output sequence's coverage over the input. It requires more complicated coverage computation based on all accumulated attention weights for each hypothesis up to the current search step and involves several threshold values to be tuned. \cite{Zoltan2019S2S} combined length reward and coverage term, and obtained stable results up to beam size 240. However, the resulting system has seven individual hyper-parameters to be optimized for decoding, which is a huge tuning effort. 

While these approaches largely eliminate the length bias problem, they are either pure heuristics or difficult to optimize. Their usage and results are mostly reported with beam sizes below 240. The potential side-effect due to the additional bias introduced towards longer sequences is often disregarded. We show that with a much larger beam size, such bias leads to more wrong decisions towards too long transcriptions, which results in severe performance degradation. This suggests that the heuristic approaches are not proper modeling refinement and make the decisions less robust w.r.t. length variation and search beam size.
 
In this work, we propose a novel beam search derived from reinterpreting the sequence posterior with an explicit length modeling. By applying the reinterpreted probability together with beam pruning, the resulting final probability is obtained from pure estimations based on models' output without heuristics. This leads to a robust model modification which allows reliable comparison among output sequences of different lengths. Experimental verification on the LibriSpeech corpus \cite{libsp} shows that the proposed approach eliminates the length bias problem without heuristics or additional tuning effort. Compared with the heuristic baseline, it achieves better performance and shows better efficiency with the additional derived early stopping criterion. More importantly, without introducing additional side-effects, the proposed approach provides robust decision making and consistently good performance under both small and very large beams. It is also applicable to streaming usage as well as other tasks such as NMT.

\vspace{-1mm}
\section{Proposed Beam Search}
\vspace{-0.5mm}
\subsection{Probability reinterpretation}
Let $x_1^T$ denote an input sequence of length $T$ and $a_1^N$ denote partial output sequence hypotheses at the $N$-th step of beam search, where $N$ also represents output label position. The original sequence posterior probability of $a_1^N$ is quantified by:\\
\scalebox{0.9}{\parbox{1.11\linewidth}{%
\begin{align*}
q(a_1^N|x_1^T) & = p(a_1^N|x_1^T) \cdot p^{\alpha}(a_1^N) \\
               & = \prod_{n=1}^N p(a_n|a_0^{n-1}, x_1^T) \cdot p^{\alpha}(a_n|a_0^{n-1}) \numberthis \label{eq:origProb}
\end{align*}}}
The optional language model (LM) shallow fusion \cite{shallowFusion2015} with scale $\alpha$ can be omitted without influencing the derivation. 

Let $V \cup \{\$\}$ define the output label vocabulary, where $\$$ is the sequence end label. If $a_N=\$$, then $a_1^N$ represents ending sequences at position $N$. Ending sequences are terminated without further expansion and are stored separately. Therefore, $a_N=\$$  also implies $a_1^{N-1} \in V^{N-1}$, which we omit in all equations for simplicity. By considering $\$$ as the last output label, the output sequence length of ending sequences $a_1^N$ is obtained as $len=N$, which reversely implies $a_N=\$$.
For ending sequences at position $N$, we rewrite their final probability with an explicit length modeling:\\
\scalebox{0.9}{\parbox{1.1\linewidth}{%
\begin{align*}
p(a_1^N, len=N| x_1^T) &= p(a_1^N| len=N, x_1^T) \cdot p(len=N | x_1^T) \\
p(a_1^N| len=N, x_1^T) &= \frac{\underset{{a_N=\$}}{q}(a_1^N|x_1^T)}{\sum_{\{\hat{a}_1^N:\hat{a}_N=\$\} \in B_N} q(\hat{a}_1^N|x_1^T)} \\
p(len=N | x_1^T) &= p_{N}(\$|x_1^T) \prod_{n=1}^{N-1} (1-p_n(\$|x_1^T)) \numberthis \label{eq:lenProb}\\
p_{N}(\$|x_1^T) &= \frac{\sum_{\{a_1^N:a_N=\$\} \in B_N} q(a_1^N|x_1^T)}{\sum_{\hat{a}_1^N \in B_N} q(\hat{a}_1^N|x_1^T)} \numberthis \label{eq:endProb}
\end{align*}}}
Here $B_N=\{a_1^N| a_1^{N-1} \in V^{N-1}, a_N \in V \cup \{\$\}\}$ is an unlimited beam of all label sequence hypotheses reaching position $N$, which can end at positions larger or equal to $N$. Additionally, we define $p_{N}(\$|x_1^T)$ as the ending probability at position $N$. It is obtained by re-normalizing the probability mass of all label sequences ending at position $N$ over the probability mass of all label sequences reaching position $N$.
Accordingly, $1-p_N(\$|x_1^T)$ accounts for the non-ending probability at position $N$. 
Therefore, the probability of finishing with output sequence length $len = N$, i.e. \Cref{eq:lenProb}, can be obtained by multiplying the accumulated non-ending probabilities from positions $1$ to $N-1$ with the ending probability at position $N$. By merging all terms, we obtain the final probability of ending sequences at position $N$ as: \\
\scalebox{0.9}{\parbox{1.11\linewidth}{%
\begin{equation}
p(a_1^N, len=N| x_1^T) = \underbrace{\frac{\underset{{a_N=\$}}{q}(a_1^N|x_1^T)}{\sum_{\hat{a}_1^N \in B_N} q(\hat{a}_1^N|x_1^T)}}_{p_B} \cdot \underbrace{\prod_{n=1}^{N-1} (1-p_n(\$|x_1^T))}_{p_{!\$}} \label{eq:finalProb}
\end{equation}}}

Note that with an unlimited beam $B$ at each step, $\prod_{n=1}^{N-1} (1-p_n(\$|x_1^T))$ is equal to $\sum_{\hat{a}_1^N \in B_N} q(\hat{a}_1^N|x_1^T)$. Both represent the probability mass of all label sequence hypotheses reaching position $N$. This verifies the derivation of the reinterpreted final probability which leads to the same sequence posterior as in \Cref{eq:origProb}. Note that no additional parameters or model training are introduced here.

\subsection{Beam search with pruning}
We then apply this reinterpreted final probability \Cref{eq:finalProb} into normal beam search, where $B_N$ becomes a limited beam after pruning. At each search step $N$, we use the sequence posterior in \Cref{eq:origProb} to directly prune all partial label sequence hypotheses $a_1^N$. Since all of them have the same length up to this position, they are directly comparable. We first apply score-based pruning to prune away hypotheses whose score difference to the current best is more than a predefined threshold. A predefined beam size is then applied if the remaining number of hypotheses still exceeds this upper bound. 

Ending sequences are then detected from the remaining hypotheses in the beam $B_N$, which are used to compute the ending probability of position $N$ according to \Cref{eq:endProb}. We apply the accumulated non-ending probability from all previous positions $1$ to $N-1$ into \Cref{eq:finalProb} to compute the final probability for each ending sequence within $B_N$. Since all computation only has a dependency on the past, no additional delay is introduced here. All ended sequences up to this search step are stored separately and we only keep the best $k$ of them based on their final probability. Note that we explicitly do not use ended sequences to prune away ongoing sequences in further steps, since they may not be directly comparable.

\subsection{Final probability}\label{endSeq}
With such limited beam at each search step, the obtained final probability no longer equals to the original sequence posterior in \Cref{eq:origProb}. It essentially leads to a beam-dependent model modification. The fraction term (denoted as $p_B$ in \Cref{eq:finalProb}) can be interpreted as renormalization within $B_N$, which estimates the relative quality of the ending sequence within the beam of the current step. The non-ending probability of each previous position is effectively also renormalization within the corresponding beam, which indicates how probable it is to not end at that position. The accumulated non-ending probability from all previous positions (denoted as $p_{!\$}$ in \Cref{eq:finalProb}) then estimates the probability of not finishing before the current position $N$. Both $p_B$ and $p_{!\$}$ are pure estimations based on the models' output without heuristics, which jointly decide the final probability of ending sequences. 

Note that this final probability depends on the beam pruning. For extremely large beams with little pruning, it approaches the original sequence posterior which may still suffer the length bias problem. For extremely small beams with very strong pruning, it can have overestimation problem and search become less reliable, which however does not contradict the concept of beam search. Both cases are very unlikely by simply applying a reasonable threshold for the score-based pruning. This leads to an optimal beam at each step based on scores, which prunes away bad hypotheses while keeping a proper probability mass for renormalization. We observe that even without score-based pruning, the approach works consistently well with both small and very large beams.

In terms of reliable comparison among ending sequences of different lengths based on this final probability, some intuitive interpretation can be given as following. Let $M_{\text{opt}}$ denote the correct output sequence length for a given input sequence. At positions much smaller than $M_{\text{opt}}$, ending sequences should have rather small sequence posterior based on a reasonable model. Even if they survive pruning, their final probability should have a high $p_{!\$}$ but suffer a very low $p_B$. At positions around $M_{\text{opt}}$, sequence posterior of ending sequences close to the correct transcription become more dominant in the beam. This leads to an increasing $p_B$ and a one-step-delayed decreasing $p_{!\$}$. These ending sequences should have a rather high final probability. Finally at positions much larger than $M_{\text{opt}}$, ending sequences might have a good $p_B$, but suffer a very low $p_{!\$}$.

\subsection{Decision and early stopping}
The final best output sequence can be decided using the maximum a posteriori (MAP) decision rule:\\
\scalebox{0.9}{\parbox{1.11\linewidth}{%
\begin{align*}
x_{1}^{T} \rightarrow \underset{\text{opt}}{a_1^M} = \argmax_{a_1^M, M} p(a_1^M, len=M| x_1^T)
\end{align*}}}
Since we do not apply pruning between ended sequences and ongoing sequences in further steps, we need to derive a stopping criterion to avoid unnecessary search steps. This can be easily obtained from \Cref{eq:finalProb}. Let $\tilde{a}^M_1$ denote the current best ended sequence, where $1 \leq M \leq N$ and $N$ is the current step. All future hypotheses from further steps after $N$ can not be better than $\tilde{a}^M_1$, if the following holds:\\
\scalebox{0.9}{\parbox{1.11\linewidth}{%
\begin{align*}
\prod_{n=1}^N (1-p_n(\$|x_1^T)) \leq p(\tilde{a}^M_1, len=M| x_1^T)
\end{align*}}}
An additional maximum length constraint with respect to the input sequence length can also be added to stop decoding, which is generally valid for ASR. The pseudo code of the proposed beam search is given in Algorithm~\ref{alg:algo}, where the choice of $a_0$ and the initial computation with or without $a_0$ can be model-specific.


\vspace{-1mm}
\section{Experiments}
\vspace{-1mm}
\subsection{Setups}
The proposed beam search is implemented based on the RWTH ASR toolkit\footnote{Source code will be published in the next release of RASR.}  \cite{rasr} with an extension described in \cite{Beck20191pass}. Experiments are conducted on the LibriSpeech corpus \cite{libsp}. Both the long short term memory \cite{Hochreiter1997lstm} (LSTM)-based encoder-decoder attention model and the LSTM LM are the same as described in \cite{Zeyer2019ASRU}. They are trained on the LibriSpeech acoustic and LM training data respectively using the RETURNN toolkit \cite{returnn, Zeyer2018returnn}. Both models share the same set of about 10k byte-pair encoding (BPE) units. We refer the readers to \cite{Zeyer2019ASRU} for more model and training details. Different beam sizes from $\{32, 64, 128, 5000\}$ are evaluated. Decoding parameters are optimized on each development set and applied to the corresponding test set. All results are obtained with the MAP decision rule.

\begin{algorithm}[t!]
\SetKwInOut{Init}{Initialize}
\SetAlgoLined
\caption{Proposed Beam Seach}\label{alg:algo}
\Init{$N=0, B_0=\{a_0\}, k_\text{best}=\{\},$ \\ $p_{{!\$}}=1.0, Stop=\text{false}$}
\While{\text{not} $Stop$}{
  $N \pluseq 1$\;
  \For{$a_1^{N-1}$ in $B_{N-1}$}{
    extend to all $a_1^N$ and compute $q(a_1^N|x_1^T)$\;
    add all $a_1^N$ to $B_N$\;
  }
  remove $B_{N-1}$\;
  apply beam pruning in $B_N$\;
  $p_{\sum} = 0, p_{\sum_\$}=0, B_{\$}=\{\}$\;
  \For{$a_1^{N}$ in $B_{N}$}{
    $p_{\sum} \pluseq q(a_1^N|x_1^T)$\;
    \If{$a_N == \$$}{
      $p_{\sum_\$} \pluseq q(a_1^N|x_1^T)$\;
      move $a_1^N$ from $B_N$ to $B_{\$}$\;
    }
  }
  \For{$a_1^{N}$ in $B_{\$}$}{
    $p_\text{final}(a_1^N, len=N| x_1^T) = q(a_1^N|x_1^T) / p_{\sum} \cdot p_{{!\$}}$\;
    insert $a_1^N$ to $k_\text{best}$ based on $p_\text{final}(a_1^N, len=N| x_1^T)$\;
  }
  $ p_{{!\$}} \prodeq (1- p_{\sum_\$} / p_{\sum})$\;
  \If{$ p_{{!\$}} \le$ best $p_\text{final}$ in $k_\text{best}$ or $N \geq T$}{
    $Stop =$ true\;
  }
}
\Return{$k_\text{best}$}
\end{algorithm}

\begin{table}[ht!]
\footnotesize
\caption{\it WER comparison of different beam search with different beam sizes on the LibriSpeech corpus. Additional analysis on the dev-other set including insertion, deletion and substitution errors, and average transcription length (BPE units are merged to words already). Reference transcriptions of dev-other set have an average length of $17.8$ words.}
\vspace{-1mm}
\setlength{\tabcolsep}{0.22em}
\begin{center}\label{tab:wers}
\begin{tabular}{|c|c|r|r|r|r|c|c|c|c|}
\hline
Beam & Beam & \multicolumn{2}{c|}{dev} & \multicolumn{2}{c|}{test} & \multicolumn{4}{c|}{dev-other} \\ \cline{3-10}
Search & Size & clean & other & clean & other & ins & del & sub & len\\ \hline
\multirow{2}{*}{\hspace{-5mm}simple} & 64 & 5.9 & 11.1 & 6.5 & 12.2 & 0.6k & 1.3k & 3.8k & 17.5 \\ \cline{2-10}
                                    & 5000 & 19.7 & 32.0 & 20.3 & 35.3 & 0.3k & 13.5k & 2.5k & 13.2 \\ \hline
\multirow{5}{*}{  + heuristics} & 32         & 2.8 & 7.6  & 3.1 & 8.3 & \multicolumn{4}{c|}{n.a.} \\ \cline{2-10}
                                & 64         & 2.8 & 7.5 & 3.1 & 8.2 & 0.6k & 0.2k & 3.0k & 17.9 \\ \cline{2-10}
                                & 128        & 2.8 & 7.7 & 3.1 & 8.7 & \multicolumn{4}{c|}{n.a.}\\ \cline{2-10}
                                & 5000       & 5.2 & 15.7 & 5.7 & 17.8 & 4.6k & 0.2k & 3.1k & 19.3 \\ \hline
\multirow{5}{*}{\hspace{-2mm}proposed} & 32   & 2.8 & 7.4 & 3.1 & 8.0 & \multicolumn{4}{c|}{n.a.} \\ \cline{2-10}
                          & 64   & 2.8 & 7.2 & 3.1 & 7.9 & 0.5k & 0.2k & 2.9k & 17.8\\ \cline{2-10}
                          & 128  & 2.8 & 7.2 & 3.1 & 7.9 & \multicolumn{4}{c|}{n.a.} \\ \cline{2-10}
                          & 5000 & 2.8 & 7.2 & 3.1 & 8.0 & 0.5k & 0.3k & 2.8k & 17.8\\ \cline{2-10}
                          & optimal & 2.8 & 7.1 & 3.1 & 7.8 & \multicolumn{4}{c|}{n.a.} \\ \hline
\end{tabular}
\end{center}
\vspace{-4mm}
\end{table}

\begin{table*}[t!]
\footnotesize
\centering
\caption{\it Example transcription with scores and average number of search steps for heuristic-based and proposed beam search with different beam sizes on the LibriSpeech dev sets (BPE units are merged to words already).}
\label{tab:example}
\setlength{\tabcolsep}{0.6em}
\begin{tabular}{|c|c|c|c|c|c|c|}
\hline
\multirow{2}{*}{Beam Search} & Beam & Example of Recognized Transcription & Original & Final & \multicolumn{2}{c|}{Search Steps} \\ \cline{6-7}
 & Size & (utterance 1585-157660-0003) & Score & Score & dev-clean & dev-other \\ \hline
\multirow{3}{*}{simple+ heuristics} & 64 & ``GLORIOUS LONDON" & -10.55 & -3.52 & 27.0 & 25.6 \\ \cline{2-7}
 & \multirow{2}{*}{5000} & ``ZARATHUSTRA DE L'OISEAU DE L'OISEAU & \multirow{2}{*}{-111.39} & \multirow{2}{*}{-3.28} & \multirow{2}{*}{48.0} & \multirow{2}{*}{49.1} \\
& & DE L'OISEAU DE L'OISEAU"&  &  & & \\ \hline
\multirow{2}{*}{proposed} & 64 & \multirow{2}{*}{``GLORIOUS LONDON"} & \multirow{2}{*}{-10.55} & -0.32 & 24.2 & 21.7 \\ \cline{2-2} \cline{5-7}
                          & 5000 & & & -0.45 & 24.2 & 21.8 \\ \hline
\end{tabular}
\vspace{-5.5mm}
\end{table*}

\subsection{Simple beam search with heuristics}\label{baseline}
We follow \cite{Zeyer2019ASRU} to apply simple beam search with heuristics using length normalization and EOS threshold. Here the scale for LM shallow fusion and the EOS threshold factor need to be optimized. \cite{Zeyer2019ASRU} reported to achieve the best result with beam size 64. We apply the optimal parameter settings for beam size 64 to all other beam sizes. For a complete comparison, we also include the results of simple beam search without heuristics under beam sizes 64 and 5000. The word error rate (WER) results are shown in \Cref{tab:wers}. Additionally for the dev-other set, we show insertion, deletion and substitution errors as well as the average length of recognized transcriptions under beam sizes 64 and 5000. The trend remains the same also for other subsets.

Without heuristics, simple beam search suffers a huge increment of deletion errors from beam size 64 to 5000. The length bias problem and corresponding beam problem are directly visible from the largely degraded performance and much shorter transcription lengths. This indicates a major flaw in modeling, which clearly requires modeling refinement. Heuristics using length normalization and EOS threshold reduce the deletion errors and improve the results dramatically. Rather stable performance is obtained with beam sizes 32, 64 and 128 except a small degradation on the test-other set. For beam size 64, the average length of recognized transcriptions closely approaches the one of the reference transcriptions ($17.8$ words), which shows a good effectiveness against the length bias problem. However, a considerable performance degradation is observed with beam size 5000. The major impact comes from a large increment of insertion errors, which is also visible from the longer transcription length. This raises a new beam problem suggesting that heuristics are not proper modeling refinement.

We also conduct informal experiments to apply separate pruning between ended and ongoing sequences, and use the input length constraint to stop decoding. This gives worse results for both simple beam search with and without heuristics.

\subsection{Proposed beam search}
For the proposed beam search, only one scale for LM shallow fusion needs to be optimized. For a fair comparison under the same beam sizes, we explicitly deactivate the score-based pruning. We optimize the LM scale for beam size 64 and apply it to all other beam sizes. The results are also shown in \Cref{tab:wers}. 

Compared with simple beam search without heuristics, the proposed approach clearly eliminates the length bias problem based on the largely reduced deletion errors and improved accuracy. Unlike the heuristic approach, this effectiveness is maintained when the beam size is increased from 64 to 5000. For both beam sizes, it produces the same average transcription length as the reference, which strongly supports the intuitive interpretation given in \Cref{endSeq} about reliable comparison among output sequences of different lengths. In fact, consistent and good performance is obtained under all beam sizes. This suggests a more robust capability for modeling refinement, even though the approach does not provide a theoretical final solution to the length bias problem. Compared with the best WER of the heuristic baseline using beam size 64, the proposed approach achieves the same performance on the `clean' sets and 4\% relative improvement on the `other' sets.  To show the performance of the complete approach, we also include results using score-based pruning with a threshold of 8 and beam size 5000 as upper bound. Further improvement is obtained on the `other' sets by using such optimal beam at each step.    
                        
\subsection{Analysis}\label{analysis}
For more insights into the new beam problem of the heuristic approach, we further check those utterances of degraded performance from beam size 64 to 5000. We find out that they actually point out a robustness issue of the heuristic-based score for decision making. For better illustration, we show one example of such utterances in \Cref{tab:example}. We denote the score of \Cref{eq:origProb} as original score and the approach-specific score as final score. 

Based on the length-normalized final score, the heuristic-based beam search decides for the correct transcription with beam size 64 and a much longer transcription with beam size 5000. However, this wrong transcription actually has a much worse original score based on the models' output. This indicates a strong bias introduced by the heuristics towards longer output sequences, which can over-correct the length bias and cause new  modeling problems. With much larger beam and more hypotheses considered, this leads to more wrong decisions towards too long transcriptions. Therefore, good performance is still only achievable with rather small beam sizes and careful tuning, where search errors are adopted to cover the new modeling errors. This completely contradicts the concept of beam search. Similar effect is also possible with other heuristics such as length reward. \cite{Bahdanau2016e2e} applied length reward in decoding and reported issues about looping transcriptions. This is very similar as the example shown here, which is actually resulted from the same reason. 

In contrast, without introducing any artificial terms, the proposed approach gives the best final score for the correct transcription under both beam sizes. This reflects the robustness of the proposed final probability for decision making, which serves as a robust model modification as described in \Cref{endSeq}. For both small and very large beam sizes, the approach solves the length bias problem without introducing additional side-effects, which also explains the performance difference in \Cref{tab:wers}.

\subsection{Efficiency}
In terms of computation at each search step under the same beam size, there is not much difference between the baseline and proposed approach. However, since we do not use ended sequences to prune away ongoing sequences, we need to check if our derived stopping criterion really avoids unnecessary search steps. We verify this by comparing the average number of search steps needed to finish recognition for each dev set under beam sizes 64 and 5000. As shown in the last two columns of \Cref{tab:example}, the numbers needed for the simple beam search with heuristics largely increase with beam sizes. Therefore, its efficiency decreases with increasing beam sizes. On the other hand, the numbers needed for the proposed approach are consistently small under both beam sizes. This approves the derived early stopping criterion and the better efficiency of the approach.

\vspace{-2mm}
\section{Conclusion}
\vspace{-0.5mm}
In this work, we presented a novel beam search derived from reinterpreting the sequence posterior with an explicit length modeling. By applying the reinterpreted probability together with beam pruning, the obtained final probability leads to a robust model modification without heuristics, which allows reliable comparison among output sequences of different lengths. Experiments on the LibriSpeech corpus show that the proposed approach solves the length bias problem without heuristics or additional tuning effort. We showed that simple heuristics are not proper modeling refinement and introduce strong bias for decision making, which results in severe performance degradation with largely increased beam size. In contrast, the proposed approach provides robust decision making and consistently good performance under both small and very large beam sizes. 
Compared with the best WER of the heuristic baseline using small beam size as in practice, the approach achieves the same performance on the `clean' sets and 4\% relative improvement on the `other' sets. 
It is also more efficient with the additional derived early stopping criterion.

Future work includes verifying the proposed approach with different data and more complicated decision rules, and extension to a more general label-synchronous search framework. It is also worthy to further research into better modeling approaches that in principle would work even without pruning and thus fully retain a proper beam search behavior. 

\vspace{-2mm}
\section{Acknowledgements}
\vspace{-0.5mm}
\footnotesize
\setstretch{0.8}
This work has received funding from the European Research Council (ERC) under the European Union's Horizon 2020 research and innovation program (grant agreement No 694537, project ``SEQCLAS") and from a Google Focused Award. The work reflects only the authors' views and none of the funding parties is responsible for any use that may be made of the information it contains.

We thank Wilfried Michel for useful discussions.

\bibliographystyle{IEEEtran}
\bibliography{ref}

\end{document}